\documentclass[conference]{IEEEtran}
 
\usepackage[english]{babel}
\usepackage[export]{adjustbox}
\usepackage[justification=centering]{caption}
\usepackage[section]{placeins}
\usepackage[T1]{fontenc}
\usepackage[table]{xcolor}
\usepackage[utf8]{inputenc}
\usepackage{algorithm,algcompatible,amsmath}
\usepackage{amsfonts}
\usepackage{amssymb}
\usepackage{balance}
\usepackage{blindtext}
\usepackage{cases}
\usepackage{cite}
\usepackage{epsfig}
\usepackage{epstopdf}
\usepackage{fancyhdr}
\usepackage{float}
\usepackage{graphicx}
\usepackage{latexsym}
\usepackage{lettrine}
\usepackage{lipsum, color}
\usepackage{makecell}
\usepackage{mathtools, cuted}
\usepackage{multicol,lipsum}
\usepackage{multirow}
\usepackage{pdfpages}
\usepackage{pifont}
\usepackage{pst-all}
\usepackage{pstricks}
\usepackage{resizegather}
\usepackage{setspace}
\usepackage{stfloats}
\usepackage{subcaption}
\usepackage{upgreek}

\algnewcommand\INPUT{\item[\textbf{Input:}]}%
\algnewcommand\OUTPUT{\item[\textbf{Output:}]}%
\algnewcommand\FIND{\item[\textbf{Find}]}
\algnewcommand\INT{\item[\textbf{Initialize}]}
\algnewcommand\REE{\item[\textbf{Repeat}]}
\algnewcommand\IFF{\item[\textbf{If}]}
\algnewcommand\SD{\item[\textbf{For}]}

\allowdisplaybreaks

\graphicspath{{myfig/}}
\graphicspath{ {./results/} }

\makeatletter

\IEEEoverridecommandlockouts
\vspace{-0.1in}

\IEEEoverridecommandlockouts
\vspace{-0.1in}
\title{ Energy Efficiency Fairness  Beamforming  Designs for MISO NOMA Systems
}
\vspace{-0.1in} 
\author{\IEEEauthorblockN{Haitham Al-Obiedollah\IEEEauthorrefmark{1}, Kanapathippillai Cumanan\IEEEauthorrefmark{1}, Jeyarajan Thiyagalingam\IEEEauthorrefmark{2},  Alister G. Burr\IEEEauthorrefmark{1},\\ Zhiguo Ding\IEEEauthorrefmark{3}, and Octavia A. Dobre\IEEEauthorrefmark{4}}\\
\IEEEauthorblockA{\IEEEauthorrefmark{1}Department of Electronic Engineering, University of York, York, YO10 5DD, UK\\
\IEEEauthorblockA{\IEEEauthorrefmark{2}STFC, Rutherford Appleton Laboratory, Oxford, OX11 0QX, UK
\\
\IEEEauthorblockA{\IEEEauthorrefmark{3} School of Electrical and Electronic Engineering, The University of Manchester,
Manchester, UK}\vspace{-0.2in} \\
\IEEEauthorblockA{\IEEEauthorrefmark{4}Department of Electrical and Computer Engineering, Memorial University, St. John’s,  Canada}\vspace{-0.2in} \\
Email: \{hma534, kanapathippillai.cumanan, alister.burr\}@york.ac.uk\IEEEauthorrefmark{1},\\ t.jeyan@stfc.ac.uk\IEEEauthorrefmark{2},  zhiguo.ding@manchester.ac.uk\IEEEauthorrefmark{3}, odobre@mun.ca\IEEEauthorrefmark{4}}
}}

\begin{document}
\maketitle
\vspace{-0.2in} 
\linespread{.94}
\begin{abstract}
In this paper, we propose two beamforming designs for a multiple-input
single-output   non-orthogonal multiple access  system
considering the energy efficiency (EE) fairness between users. In
particular, two quantitative fairness-based designs are developed to
maintain fairness between the users in terms of achieved EE: max-min
energy efficiency (MMEE) and proportional fairness (PF) designs. While
the MMEE-based design aims to maximize the minimum EE of the users in
the system, the PF-based design aims to seek a good balance between
the global energy efficiency   of the system and the EE fairness
between the users.  Detailed simulation results indicate that our
proposed designs offer many-fold EE improvements over the existing
energy-efficient beamforming designs.
\end{abstract}
\IEEEpeerreviewmaketitle

\section{Introduction}

Non-orthogonal multiple access (NOMA) has been recently envisioned as
a promising multiple access technique to be used in future wireless
networks for addressing the issue of low spectral efficiency in
conventional orthogonal multiple access (OMA) and to provide massive
connectivity~\cite{fnoma}. In this novel multiple access scheme,
multiple users share the same orthogonal resources (i.e., time,
frequency and spreading codes) by exploiting power-domain
multiplexing~\cite{noma2}.  In particular, superposition coding (SC)
is employed at the transmitter to multiplex the signals of multiple
users in the power domain~\cite{noma2}, and then,  the successive
interference cancellation (SIC) technique is used at the receiving
ends to eliminate inter-user interference and to decode the signals \cite{noma4,fayzeh}.

To facilitate a practical implementation of NOMA in dense networks,
and to further improve the spectral efficiency \cite{misonoma}, NOMA
is incorporated with multiple antenna techniques to exploit their
additional degrees of freedom offered by the different spatial
multiplexing schemes \cite{mimonoma} \cite{cuma8}. As such, NOMA has the
potential capabilities to support the extensive deployment of the
Internet-of-Things (IoT) in fifth generation (5G) and beyond wireless
networks \cite{noma4}. However, the limitation of the available power
resources becomes  one of the major challenges in the development of
future technologies. This should be taken into account in the design
of new transmission techniques \cite{book}.  The energy efficiency
(EE),   defined as the ratio between the
achieved sum rate in the system and the total power consumption to
achieve those rates at users \cite{haitham}, is a useful metric for
comparing and characterizing different schemes, such as beamforming
designs.  Furthermore, the EE can also help to strike a good balance
between the achieved rate in the system and the total power
consumption \cite{ee1}. Note that the terms EE and global energy efficiency (GEE) carry the same meaning in this paper.   In particular, GEE considers the overall EE of the system without taking the performance of the individual users into
account.  Hence, the users with weaker channel conditions (i.e.,
cell-edge users) might achieve a very low EE compared to those users
with stronger channel conditions (near users). To overcome such a
fairness issue among the users, the transmitter should be able to
incorporate the performance of the individual users in the design
rather than optimizing the GEE of the system. Furthermore, while there is no
unique definition for fairness, this could be generally
defined in terms of allocating the resources between the users to
provide a reasonable quality-of-service at all of
them \cite{fairness}.

Motivated by the prominence of the fairness in terms of the achieved
EE for each user, we consider energy-efficient fairness-based
beamforming designs for a multiple-input single-output (MISO) NOMA
system. {  The beamforming design with GEE   is considered for a MISO NOMA system   in  \cite{haitham}.}
In particular, we present two fairness based designs in this
paper, namely, max-min energy efficiency (MMEE) and proportional
fairness (PF) based designs. First, MMEE is considered as the
bottleneck fairness design \cite{pricefair}  \cite{cuma10}.  As such, MMEE is
achieved if any performance increment in the EE of the $i^{th}$ user
(EE$_i$) causes a deterioration of the EE of the $j^{th}$ user (i.e.,
EE$_j$) which already has lower performance \cite{maxmin}.  Despite the fact that
the MMEE design aims to achieve the same EE for all users by
maximizing the minimum EE of a user, the fairness in this design comes
at the cost of GEE degradation. Therefore, we consider another
approach, namely  the PF-based design, which has the  capability
to finding a good balance between  GEE and  achieved EE for
each user \cite{pricefair}.  Assume that a design achieves an EE of EE$_i$ at the $i^{th}$ user in
a system with $K$ users by allocating an amount of $\Gamma_{i}$
resources.   Then, the resource   allocation   $\{\Gamma^*_i\}_{i=1}^{K} $  is considered to be a  \textit{proportionally  fair }if   the following condition holds for any other feasible resource allocation $\{\Gamma_i\}_{i=1}^{K} $    \cite{fairness}:
\begin{equation}\label{pfq}
 \sum_{i=1}^{K} \frac{\text{EE} _i-\text{EE}^* _i}{\text{EE}^* _i} \leq 0, 
\end{equation} 
where $\text{EE}_i^*$ corresponds to $\Gamma^*_i$. It is worth mentioning that the condition in (\ref{pfq}) can be
satisfied through determining the feasible set
$\{\Gamma^*_i\}_{i=1}^{K} $ that maximizes $\sum_{i=1}^{K}\log
(\text{EE} _i) $ \cite{ fairmain}.
In this paper, we formulate fairness-based beamforming designs (i.e.,
MMEE and PF) for a MISO NOMA system with total power constraint at the
base station (BS) and minimum rate requirement at each user. However,
these optimization problems are non-convex in nature in terms of
beamforming vectors. Hence, we employ the sequential convex
approximation (SCA) technique to tackle the non-convexity issues
associated with these optimization problems. In addition, we demonstrate
the effectiveness of the proposed designs by evaluating and comparing
their performances with that of the existing beamforming designs in
the literature. 
 
The rest of the paper is organized as follows. In Section~\ref{sec2},
the system model and   problem formulations are
introduced. Section~\ref{sec3} presents the SCA technique as an effective
approach to solve the original non-convex optimization problems. To
validate the performance of the proposed designs, numerical results
are provided in Section~\ref{sec4}. Finally,   conclusions of this
work are presented in Section~\ref{sec5}.

\subsubsection*{Notations}

We use lower case boldface letters for vectors and upper case boldface
letters for matrices. $(\cdot )^H $ denotes complex conjugate
transpose.  $\Re(\cdot)$ and $\Im(\cdot)$ stand for real and imaginary
parts of a complex number, respectively. The symbols $\mathbb{C}^{N}$
and $\mathbb{R}^{N}$ denote $N$-dimensional complex  and real space,
respectively.  $||\cdot||_2 $ and $| \cdot| $ represent the Euclidean
norm of a vector and absolute value of a complex number, respectively.

\section{System Model And Problem Formulations}\label{sec2}
\vspace{-.09 in} 
\subsection{System Model}

We consider a downlink transmission of a MISO NOMA system, in which 
a BS equipped with $N$ multiple antennas communicates simultaneously
with $K$ single-antenna users. {  It is assumed that BS has the perfect channel state information of each user}. The BS encodes the message of the
$i^{th}$ user ($s_i$) by scaling the message using linear precoding
vector (beamforming vector) $\mathbf{w}_i$ $ \in$
$\mathbb{C}^{N\times1}$.  Thus, the transmitted signal $\mathbf{x} \in
\mathbb{C}^{N\times1}$ from the BS can be written as
\begin{equation}\label{sasra}
\mathbf{x}=\sum_{i=1 }^{K} \mathbf{w}_i s_i.
\end{equation}                                      
The received  signal at the $ i^{th}$ user   ($U_i)$ can be expressed  as  
\begin{equation}\label{sara}
y_i=\sum_{j=1 }^{K}\mathbf{h}_i^H \mathbf{w}_j s_j +n_i,    \forall i\in \mathcal{K}\overset{\bigtriangleup}{ =}\{1,2,3,\cdots,K\},
\end{equation}
where $\mathbf{h}_i$ $ \in$ $\mathbb{C}^{N\times1}$ represents the
channel vector between $U_i$ and   BS. The channel coefficients
are modeled as $ \mathbf{h}_i=
\sqrt{d_i^{-\kappa}}\mathbf{v}_i$, where $\kappa$ and $d_i$ are the
path loss exponent and the distance between $U_i$ and   BS in meter,
respectively. Furthermore, $\mathbf{v}_i$ and $n_i $ represent the
small scale fading and the zero-mean circularly symmetric complex
additive white Gaussian noise with variance $\sigma_i^2$,
respectively.
In downlink transmission of a NOMA system, the stronger users employ
SIC by firstly decoding and eliminating the interference from the
signals of the users with weaker channel conditions, and then  detect
their own signals~\cite{noma2}. Here, we assume that $U_1$ is the
strongest user with the most favourable channel condition, whereas
$U_K$ is the weakest user. In particular, the users are ordered based
on their channel strengths such that
\begin{equation}
\vert\vert \mathbf{h}_1 \vert\vert^2 \geq \vert\vert \mathbf{h}_{2} \vert\vert^2\geq \cdots\geq\vert\vert \mathbf{h}_K\vert \vert^2. 
\end{equation}
Based on this user ordering, the received signal at $U_i $ after
successfully eliminating the interference from the weaker users
through SIC can be written as
\begin{equation}\label{sara}
\overset{\sim}{y_i}=\mathbf{h}_i^H\mathbf{w}_is_i+\sum_{j=1 }^{i-1} \mathbf{h}_i^H \mathbf{w}_j s_j +n_i, \forall i\in \mathcal{K}.
\end{equation}
In particular, the message intended for the $i^{th}$ user is decoded
at the stronger user $U_k$ (i.e., $k\leq i$) with the following signal-to-interference plus noise ratio (SINR):
\begin{equation} 
\text{SINR}^{(i)}_k= \frac{\vert \mathbf{h}_k^H\mathbf{w}_i \vert^2}{\sum_{j=1 }^{i-1}\vert \mathbf{h}_k^H \mathbf{w}_j \vert^2+\sigma_k^2},   \quad \forall i\in \mathcal{K},k\leq i. 
\end{equation}
Note that  the signal intended for $U_i$ could not be correctly
decoded unless the SINR of the corresponding signal is larger than a
certain threshold ($\eta_{i}^{min}$). This explicitly means that the
SINR of decoding the $i^{th}$ user signal at the stronger users should
be greater than this threshold (i.e., $\eta_{i}^{min}$). Therefore,
the SINR of the $i^{th}$ user can be defined as in \cite{hanif}
\begin{equation}\label{sinr}
\text{SINR}_i= \text{minimum} ({\text{SINR}^{(i)}_{1}, \cdots,\text{SINR}^{(i)}_{i}}),\forall i\in \mathcal{K}.
\end{equation}
Hence, the achieved rate at $U_i$ can be expressed as 
\begin{equation}\label{rate}
R_i=  B_w\log_2 (1+\text{SINR}_i) \forall i\in \mathcal{K},
\end{equation}
where $B_w$ denotes the available bandwidth for the transmission. For
notational simplicity, we select $B_w$ to be 1.  To ensure that the
power assigned to each user in the system is inversely proportional to
its channel strengths \cite{noma3}, and to allow SIC to be
successfully implemented at the strong users \cite{hanif}, the
following conditions should be considered in the design:
\begin{equation}\label{channels}
 \vert \mathbf{h}_i^H\mathbf{w}_K \vert^2 \geq   \cdots \geq\vert \mathbf{h}_i^H \mathbf{w}_{1}\vert^2, \forall i\in \mathcal{K}.
\end{equation}
In this MISO NOMA system, the achieved EE at the $i^{th}$ user (i.e.,
EE$_i $) is defined as the ratio between the achieved rate at $U_i $
and the consumed power at the BS to achieve this rate \cite{ee1}, which
can be expressed as
\begin{equation}
 \text{EE}_i=\frac{R_i}{\frac{1}{\epsilon_0}P_i+P_{l,i}}, \forall i\in \mathcal{K},
\end{equation}
where $P_{i}$ and  $P_{l,i}$  denote    the  transmit power allocated to    $U_i $    and the corresponding  power losses associated with that user    at the BS, respectively. Note that    $P_{i}=||\mathbf{w}_i||_2^2,$  
and $\epsilon_0$ represents the efficiency of the amplifiers at the BS.  Furthermore, the  GEE of the system can  be  defined as 
\begin{equation}\label{gee}
\text{GEE}=\frac{\sum_{j=1}^{K}R_j}{\frac{1}{\epsilon_0}P +P_{l}},
\end{equation}
where $P_l$ represents the total power losses at the BS and $P$
denotes the total transmit power required for data transmission from
the BS. The available total transmit power at the BS is limited to
$P_{ava}$, which can be represented by the following constraint:
\begin{equation}\label{power1}
P=\sum_{i=1}^{K} P_i \leq P_{ava}. 
\end{equation}
In the conventional GEE maximization (GEE-Max)-based design, the
beamforming vectors are determined by maximizing the GEE under the SIC
constraints in (\ref{channels}), and with a minimum rate requirement
at each user ($ R_i^{min}$). This minimum rate requirement at the $i^{th}$
user can be imposed by the following constraint:
\begin{equation}\label{mon}
R_i \geq R_i^{min}, \forall i\in \mathcal{K}.
\end{equation} 
The  GEE-Max problem can be formulated as  
\begin{subequations}\label{Gee}
\begin{align}
 OP_{GEE}\!:
 & \underset{\{\mathbf{w}_i\}_{i=1}^{K}}{\text{max }}
& \!\!\!\!\!\!\!\!\!\!\!\!\!\! \text{GEE}\\
& ~~\text{s.t.}
& & \!\!\!\!\!\!\!\!\!\!\!\!\!\! \!\!\!\!\!\!\!\!\!\!\!   (\ref{channels}),(\ref{power1}),(\ref{mon}). \label{GE2} 
\end{align}
\end{subequations}
This GEE-Max problem $OP_{GEE}$ is solved in our previous
work using the SCA technique and the Dinkelbach's
algorithm ~\cite{haitham}. In this paper, we consider the fairness-based beamforming
designs, which are discussed in detail in the following subsection.

\subsection{Problem Formulations}

In this subsection, two fairness-based beamforming designs are
proposed, namely MMEE and PF designs.

\subsubsection{Max-min energy efficiency (MMEE)} 
 Unlike the GEE-Max-based design in $OP_{GEE}$, the MMEE design aims
for maximizing the minimum EE of users in the system while satisfying
the associated constraints \cite{cuma9}. In particular, the MMEE achieves its ideal
solution while all the users experience the same EE. However, this is
not a universal condition owing to the minimum rate and SIC
constraints. The MMEE-based design for the MISO NOMA system is
formulated as  follows:
\vspace{-0.085 in} 
\begin{subequations}\label{ee_max}
\begin{align}
 OP_{1}:
 & \underset{\{\mathbf{w}_i\}_{i=1}^{K}}{\text{max}}
& &\text{min~}\{ \text{EE}_1,\text{EE}_2,\cdots,\text{EE}_K\}\label{trt} \\
&~~ \text{s.t.}
& &   (\ref{channels}),(\ref{power1}),(\ref{mon}).
\end{align}
\end{subequations}
This max-min problem is not convex due to the non-convex objective
function in (\ref{trt}), the SIC constraint in (\ref{channels}), and
the minimum rate requirement in (\ref{mon}). Therefore, the solution
to the problem $OP_{1}$ cannot easily be determined through existing
convex optimization techniques~\cite{CVX}.
 \indent \subsubsection{Proportional Fairness (PF)} Next, we consider a
PF-based design to seek a good balance between the beamforming designs
without any fairness and with ideal fairness conditions, namely
GEE-Max design and MMEE design, respectively. The PF design can be
defined into the following optimization framework~\cite{fairness}:
\begin{subequations}\label{ee_pf}
\begin{align}
OP_{2}:
&\underset{\{\mathbf{w}_i\}_{i=1}^{K}}{\text{max}}
& &\!\!\!\!\!\!\!\!\!\!\!\!\!\!\!\!\!\!\!\sum_{i=1}^{K}\log (\text{EE}_i)\\
& ~~\text{s.t.}
& &\!\!\!\!\!\!\!\!\!\!\!\!\!\!\!\!\!\!\!(\ref{channels}),(\ref{power1}),(\ref{mon}).\label{subsub}
\end{align}
\end{subequations}
The solutions for the  non-convex problems $OP_{1}$ and
$OP_{2}$ are presented in the following section.

 \section{Proposed methodology} \label{sec3}

In this section, we exploit different techniques to convert the
original non-convex functions in $OP_{1}$ and $OP_{2}$ to convex
ones. In particular, the SCA technique is used to approximate those functions into linear convex functions~\cite{haitham,ee2,sqa}. In the
SCA technique, a set of convex lower bounds are defined with a number
of slack variables to approximate the non-convex objective function or
constraint~\cite{sqa} into a convex objective function. As such, the
SCA will be implemented to handle the non-convexity of (\ref{ee_max})
and (\ref{ee_pf}).

\subsubsection*{ Non-convex constraints in  $OP_{1}$ and $OP_{2}$} 
As both $OP_{1}$ and $OP_{2}$ share the same constraints, we first
show   how to formulate the non-convex constraints in
(\ref{ee_max}) and (\ref{ee_pf}). Without loss of generality, the
minimum rate constraint in (\ref{mon}) can be equivalently expressed in
terms of SINR as
\begin{equation}\label{soc}
\frac{\vert \mathbf{h}_k^H\mathbf{w}_i \vert^2}{\sum_{j=1 }^{i-1}\vert \mathbf{h}_k^H \mathbf{w}_j \vert^2+\sigma_k^2}\geq \eta_{i}^{min},  i\in \mathcal{K}, k\leq i,
\end{equation} 
where $\eta_{i}^{min}=2^{R_{i}^{min}}-1 $. Furthermore, this
constraint can be easily reformulated into a second order cone (SOC)
as  \cite{cvx2}:
\begin{multline}\label{const112}
\frac{1}{\sqrt{\eta_{i}^{min}}} \Re( \mathbf{h}_k^H\mathbf{w}_i )\geq\\ \vert \vert [ \mathbf{h}_k^H \mathbf{w}_{1}~ \mathbf{h}_k^H \mathbf{w}_{2}~\cdots~\mathbf{h}_k^H \mathbf{w}_{i-1}~\sigma_k]^T \vert \vert_2 ,   i\in \mathcal{K}, k\leq i.
\end{multline}
Next, the non-convexity of the SIC constraint in (\ref{channels}) is
handled by using minorization-maximization algorithm
(MMA)~\cite{hanif} \cite{wcnc2}. In particular, the non-convex function is
approximated by linear terms at a given set of values using
convex-concave procedure. Furthermore, we use the first-order Taylor
series expansion to approximate the constraint in
(\ref{channels}), where each term in the inequality is
replaced by a lower bounded linear function  $ f_{k,j}$  such that
$\vert \mathbf{h}_k^H\mathbf{w}_j \vert^2 \geq f_{k,j}$, where
\begin{multline}\label{noora1}
f_{k,j} =  |\vert [ \Re{(\mathbf{h}_k^H \mathbf{w}_j^{(n )})}~ \Im{(\mathbf{h}_k^H \mathbf{w}_j^{(n )})}]||^2_2\\+2 [ \Re{(\mathbf{h}_k^H \mathbf{w}_j^{(n )})} ~ \Im{(\mathbf{h}_k^H \mathbf{w}_j^{(n )})}]^T[(\Re{(\mathbf{h}_k^H \mathbf{w}_j )}\\-\Re{(\mathbf{h}_k^H \mathbf{w}_j^{(n )})})~(\Im{(\mathbf{h}_k^H \mathbf{w}_j )}-\Im{(\mathbf{h}_k^H \mathbf{w}_j^{(n )})})]^T,
\end{multline} 
where $\mathbf{w}_i^{(n)}$ represents the approximation of
$\mathbf{w}_i$ at the $n^{th}$ iteration. Note that the function in
(\ref{noora1}) is linear in terms of $\mathbf{w}_i $. Based on this
approximation, the non-convex constraint in (\ref{channels}) can be
approximated as the following convex constraint:
\begin{equation}\label{channeld}
 f_{i,K} \geq \cdots \geq f_{i,1}, \forall i\in \mathcal{K}.
\end{equation}
 To this end, we have approximated the non-convex constraints in the
original optimization problems $OP_{1}$ and $OP_{2}$ by convex
constraints.
\subsubsection*{MMEE Design}
In the following, we transform the original non-convex objective
function of the MMEE design in $OP_{1}$ by introducing a new slack
variable $\alpha$ as
 $$ \text{min~} \{ \text{EE}_1,\text{EE}_2,\cdots,\text{EE}_K\} \geq \alpha.$$
 Without loss of generality, the optimization problem in $OP_{1}$ can be  equivalently written as 
\begin{subequations}\label{ee_max_slack}
\begin{align}
\overset{\sim}{OP_{1}}\!\!: &
\underset{\{\mathbf{w}_i\}_{i=1}^{K}}{\text{max }} & & \!\!\!\!\!
\!\!\!\!\! \!\!\!\!\! \alpha \\ &~~ \text{s.t.}  & & \!\!\!\!\!
\!\!\!\!\! \!\!\!\!\! \!\!
(\ref{channeld}),(\ref{power1}),(\ref{const112}), \label{kog}\\ & &&
\!\!\!\!\! \!\!\!\!\! \!\!\!\!\! \!\!  \text{EE}_i \geq \alpha, i\in
\mathcal{K}.\label{ee_co1}
\end{align}
\end{subequations}
   To handle the non-convexity of the constraint in (\ref{ee_co1}), we
re-formulate this with a new slack variable $ \beta_i$ into two sets of
constraints as follows:
\begin{subequations}
\begin{align}
&R_i  \geq {\alpha \beta_i^2},\forall i\in \mathcal{K}, k\leq i,\label{eqn1s1}\\ 
&\frac{1}{\epsilon_0} \vert \vert \mathbf{w}_i \vert \vert ^2_2 + P_{l,i} \leq  {\beta_i^2},\forall i\in \mathcal{K}. \label{eqnd12}
\end{align}
\end{subequations}    
Following a similar formulation as in (\ref{const112}), the constraint
in (\ref{eqnd12}) can be cast as the following standard convex SOC:
\begin{equation}\label{fg}
 \beta_i \geq || [\frac{\mathbf{w}_i}{\sqrt{\epsilon_0}}~\sqrt{P_{l,i}}]^T||_2,\quad \forall i\in \mathcal{K}.
\end{equation} 
Furthermore, we introduce a new set of slack variables $\delta_{i}$ and $\tau_i$ to approximate the non-convex constraint in (\ref{eqn1s1}) as  
\begin{subequations}
\begin{align} 
& \log(1+\text{SINR}_k^{(i)}) \geq  \delta_i ,\forall i \in \mathcal{K}, k\leq i,\\
&  (1+\text{SINR}_k^{(i)}) \geq \tau_i, \forall i \in \mathcal{K}, k\leq i,
\end{align}
\end{subequations}   
which can be equivalently represented as the following set of
constraints:
 \begin{subnumcases}{(\ref{eqn1s1})  \Leftrightarrow  } 
    \frac{\vert \mathbf{h}_k^H\mathbf{w}_i \vert^2}{\sum_{j=1 }^{i-1}\vert \mathbf{h}_k^H \mathbf{w}_j \vert^2+\sigma_k^2}\geq \tau_i-1,  k\leq i, \label{fst}
   \\
   \tau_i \geq 2^{\delta_i}, \quad \forall i \in \mathcal{K} \label{sec}
   \\
   \delta_i \geq {\alpha \beta_i^2}, i \in \mathcal{K}.  \label{thd}
\end{subnumcases}
The non-convexity of the constraint in (\ref{fst}) is handled by
incorporating a new slack variable $\rho_{i,k}^2$ and splitting it
into the following two sets of constraints:
\begin{subequations}
\begin{align} 
&\vert \mathbf{h}_k^H\mathbf{w}_i \vert^2 \geq (\tau_i-1) \rho_{i,k}^2, \forall i \in \mathcal{K}, k\leq i,\label{huda}\\
& \rho_{i,k}^2 \geq {\sum_{j=1 }^{i-1}\vert \mathbf{h}_k^H \mathbf{w}_j \vert^2+\sigma_k^2}, \forall i \in \mathcal{K}, k\leq i\label{oof}.
\end{align}
\end{subequations}   
Following the same approach in (\ref{const112}), the constraint in
(\ref{oof}) can be transformed into a standard convex SOC constraint as
\begin{multline}\label{dsd}
\rho_{i,k} \geq \vert \vert [ \mathbf{h}_k^H \mathbf{w}_{1}~\mathbf{h}_k^H \mathbf{w}_{2}~\cdots~\mathbf{h}_k^H \mathbf{w}_{i-1}~\sigma_k]^T  \vert \vert_2, \\ \forall i \in \mathcal{K}, k\leq i. 
\end{multline}  
Furthermore,   the constraint in  (\ref{huda}) can be represented as the following convex constraint:
\begin{multline}\label{cuma}
\Re({\mathbf{h}_k^H\mathbf{w}_i})\geq\\ \sqrt{(\tau_i^{{(n) }}-1)}\rho_{i,k}^{{(n) }}+0.5\frac{1}{\sqrt{(\tau_i^{{(n) }}-1)}}\rho_{i,k}^{{(n) }}(\tau_i-\tau_i^{{(n) }})+\\  \sqrt{(\tau_i^{{(n) }}-1)}(\rho_{i,k}-\rho_{i,k}^{{(n) }}), \forall i \in \mathcal{K}, k\leq i.   
\end{multline}
Finally, we employ the first-order Taylor series expansion to approximate the right hand-side of  (\ref{thd}) as follows:
\begin{multline}\label{najw}
\delta_i \geq {\alpha^{(n)} (\beta_i^2)^{(n)}}+  (\beta_i^2)^{(n)}(\alpha-\alpha^{(n)})\\ +2 \beta_i^{(n)} \alpha^{(n)} (\beta_i-\beta_i^{(n)}), \quad i \in \mathcal{K}.
\end{multline}
After introducing these multiple slack variables, the original
non-convex MMEE optimization problem $OP_1$ is approximated as the
following optimization problem:
\begin{subequations}\label{ee_max_slacsk}
\begin{align}
\overset{\approx}{OP_1}\!\!\!:~~
& \underset{ \chi}{\text{max  }}
&  &  \!\!\!\!\! \!\!\!\!\! \!\!\!   \alpha  \\
& ~\text{s.t.}
&   &  \!\!\!\!\! \!\!\!\!\! \!\!\! (\ref{kog}),\\
&   &&  \!\!\!\!\! \!\!\!\!\! \!\!\! (\ref{fg}),(\ref{sec}),(\ref{dsd}),(\ref{cuma}),(\ref{najw}),\label{alam}
\end{align}
\end{subequations}  
where $\chi $ includes all   the  optimization variables   involved in the MMEE  problem:   $\chi  \overset{\bigtriangleup}{=}\{\mathbf{w}_k,\rho_{i,k},\tau_i,\beta_i,\delta_i,\alpha\}_{i=1}^{K}$. 
\subsubsection*{PF Design} Next, we  consider the PF maximization problem $ OP_2$. The non-convex constraints in  $ OP_2$ have been already reformulated as convex constraints in previous subsection. However,  the non-convexity of the  objective function in $ OP_2$ can be tackled by   introducing new   slack variables $\mu_i$  and  $\varsigma_i $   as
\begin{subequations}
\begin{align}
\log \text{EE}_i\geq {\varsigma_i}, \forall   i \in \mathcal{K},\\
\text{EE}_i \geq \mu_i, \forall   i \in \mathcal{K}.
\end{align}
\end{subequations}
With these new slack variables, $OP_2$ can be equivalently expressed as
\begin{subequations}\label{ee_PF1_slacsk}
\begin{align}
 \overset{\sim}{OP_2}\!\!:
 & \underset{\{\mathbf{w}_i\}_{i=1}^{K}}{\text{max }}
& & \!\!\!\!\! \!\!\!\!\! \!\!\!\!\! \!\!\!\!\! \sum_{i=1}^{K} \varsigma_i \\
& ~~\text{s.t.}
& &\!\!\!\!\! \!\!\!\!\! \!\!\!\!\! \!\!\!\!\! \mu_i \geq 2^{\varsigma_i },  i \in \mathcal{K},\label{wew}  \\
& &&\!\!\!\!\! \!\!\!\!\! \!\!\!\!\! \!\!\!\!\! \text{EE}_i\geq \mu_i,  i \in \mathcal{K}, \label{ks}\\
& && \!\!\!\!\! \!\!\!\!\! \!\!\!\!\! \!\!\!\!\!   (\ref{kog}).\label{ol}
\end{align}
\end{subequations} 
Without loss of generality, we can convert the non-convex constraint
in (\ref{ks}) to a convex one by using the same approach as in
(\ref{ee_co1}). This could be implemented by replacing $ \alpha $ in
(\ref{ee_co1}) by $\mu_i$, and then applying the corresponding
approximations.   Hence, the problem $\overset{\sim}{OP_2}$ can be
written in a convex form as
\begin{subequations}\label{ee_PF_slacsk}
\begin{align}
\overset{\approx}{OP_2}\!\!:~~ 
&\underset{\varphi}{\text{max}}
& &     \sum_{i=1}^{K} \varsigma_i \\
& ~~\text{s.t.}
& & (\ref{kog}), \\
& && \mu_i\geq 2^{\varsigma_i},  i \in \mathcal{K}, \\
& && (\ref{alam})\label{mani}, 
\end{align}
\end{subequations} 
where $\varphi$ consists of all the optimization variables: $\varphi
\overset{\bigtriangleup} {=} \{\varsigma_i,\mathbf{w}_i
,\rho_{i,k},\tau_i,\beta_i,\delta_i,\mu_i\}_{i=1}^{K}$. Note that
$\alpha$ is replaced by $ \mu_i$ at all constraints in
(\ref{mani}). \\
It is worth noting that the solutions of $\overset{\approx}{OP_1}$ and
$\overset{\approx}{OP_2}$ depend on the appropriate selection of the
initial parameters: $ \chi^{(0)} $ and $\varphi^{(0)}$.  These initial
parameters are chosen by determining the beamforming vectors
($\{\mathbf{w}_i^{(0)}\}_{i=1}^{K}$) that minimize the total transmit
power (i.e., $P=\sum_{i=1}^{K} ||\mathbf{w}_i||_2^2$) subject to the
minimum SINR constraint in (\ref{mon}) and the SIC constraint in
(\ref{channels}) \cite{cuma11}. Then, all initial parameters (i.e., $ \chi^{(0)} $
and $\varphi^{(0)}$) are evaluated by replacing the inequality with
equality at each constraint.

On the other hand, it is obvious that the solutions of
$\overset{\approx}{OP_1}$ and $\overset{\approx}{OP_2}$ can be
iteratively obtained. This iterative approach can be terminated by
comparing the difference of the objective values at two successive
iterations against a predefined threshold $\varepsilon$. We summarize
the developed algorithms  to determine the solutions of the
original MMEE and PF designs in Algorithm~1 and Algorithm~2,
respectively.

\begin{figure}[H]
    \hspace{0em}\hrulefill

 \hspace{0em} {\bf Algorithm 1:} MMEE   design  using SCA

     \hspace{0em}\hrulefill

     \hspace{0em} Step 1: Initialization of  $\chi^{(0)}$

     \hspace{0em} Step 2: Repeat
\begin{enumerate}
   \item Solve the  optimization problem $\overset{\approx}{OP_1}$ in (\ref{ee_max_slacsk}).
   \item Update $ \chi^{(n)} $ .
\end{enumerate}
\hspace{-0em} Step 3: Until required accuracy is achieved.

\hspace{-1em} \hrulefill
\end{figure} 

\begin{figure}[H]
    \hspace{0em}\hrulefill

 \hspace{0em} {\bf Algorithm 2:}  PF design  using SCA

     \hspace{0em}\hrulefill

     \hspace{0em} Step 1: Initialization of  $\varphi^{(0)}$

     \hspace{0em} Step 2: Repeat
\begin{enumerate}
   \item Solve the  optimization problem $\overset{\approx}{OP_2}$ in (\ref{ee_PF_slacsk}).
   \item Update $ \varphi^{(n)} $ .
\end{enumerate}
\hspace{-0em} Step 3: Until required accuracy is achieved.

\hspace{-1em} \hrulefill
\end{figure} 
  \section{Numerical Results} \label{sec4}
To demonstrate the effectiveness of the two proposed designs, namely
the MMEE- and PF-based designs, we perform a number of detailed
simulations. In presenting the results, we use the conventional
GEE-Max-based design as   baseline.    In our simulation studies, we
consider a downlink transmission where a BS equipped with three transmit
antennas ($N=3$)  sends signals to three users ($K=3$). It is assumed that the users are located at the distance of 1, 5.5, and 25 meters from the BS, respectively.
The
relevant parameters of the simulation setup are shown in
Table~\ref{tab:params}.
\begin{table}[tbh]
\caption{Parameters used in   simulations.}
\label{tab:params}
\begin{tabular}{|l|l|l|}
\hline
 {\bf Param.} &  {\bf Description} & {\bf Value(s)} \\
\hline
 
 $\kappa$ & Path loss exponent & $2.0$ \\
 $\sigma_i^2$    & Noise variance for user $i$ & $2.0$\\ 
 $\eta_i^{min}$  &  SINR threshold & $10^{-3}$ \\
 $P_{l,i}$       & Power loss at the BS & $45$ dBm \\
 $\epsilon_0$    &  Amplifier efficiency at   BS& $0.65$ \\
 $B_w$           & Available bandwidth         & $1$ MHz \\
 $\varepsilon$   & Thresholds for the algorithms      & $0.001$\\
 $\mathbf{v}_i$   & Small scale fading     & Rayleigh  fading \\
\hline
\end{tabular}
\end{table}
In addition  to these parameter settings, we define the normalized
transmit power (TX-SNR) in dB as TX-SNR (dB)= $10
{\log}_{10}\frac{P_{ava}}{\sigma_i^2}$. It is worth mentioning that all
simulations in this section are carried out using the CVX toolbox.

\begin{figure}[h]\label{fig:sroor}
\includegraphics[scale=.23,center]{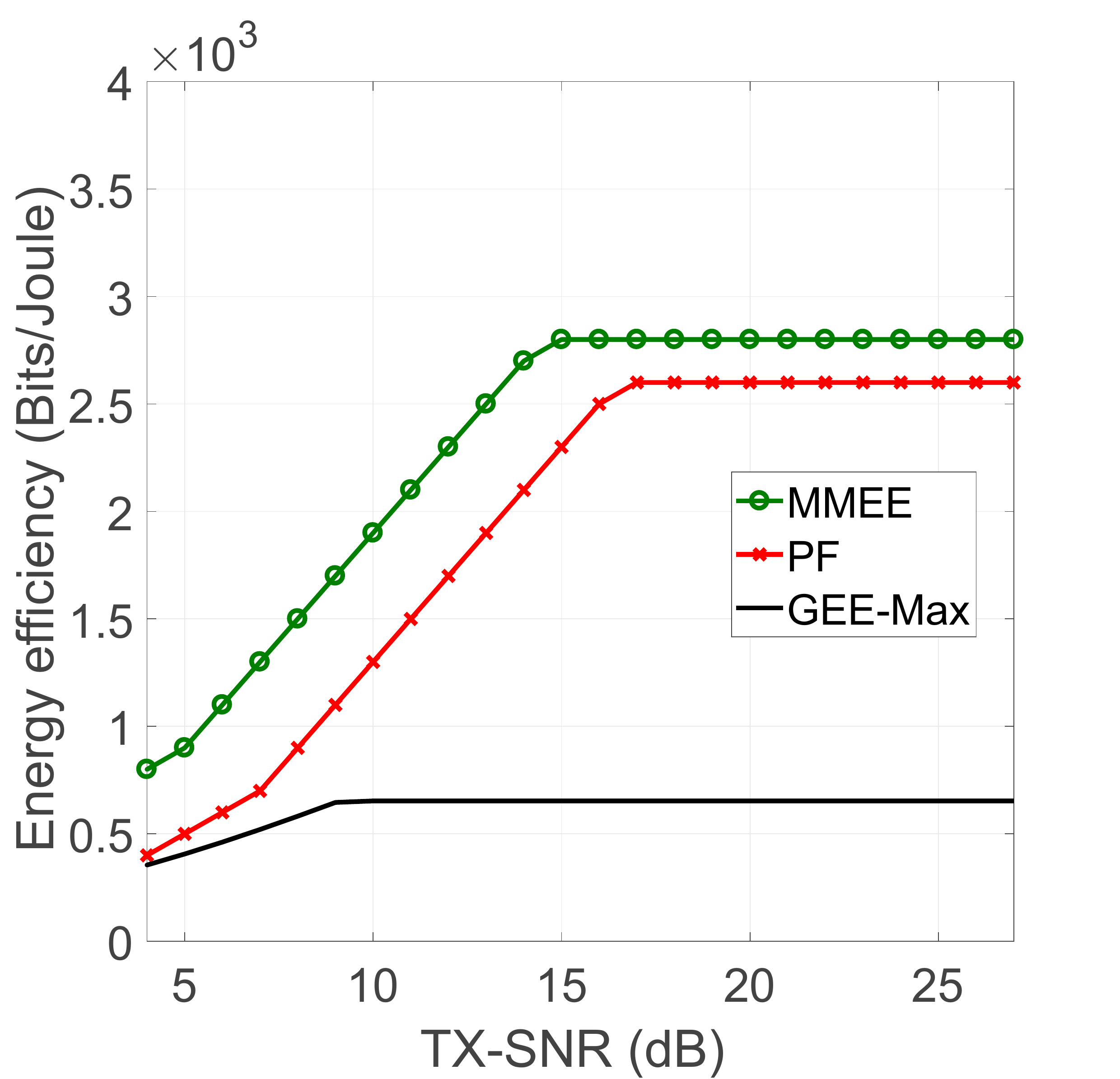}
\caption{ The performance of the weakest user with different beamforming designs.}
\centering
\label{fig:mumisoss1}
\end{figure}
Firstly, in Fig.~\ref{fig:mumisoss1} we present the achieved EE of the
weakest user in the system with different beamforming designs, namely 
GEE-Max, PF, and MMEE designs.  As can be seen in
Fig.~\ref{fig:mumisoss1}, the performance of the weakest user is
significantly improved in terms of EE when considering the MMEE and
PF-based designs compared to the conventional GEE-Max-based
design. For example, at TX-SNR=20 dB, the weakest user experiences an
EE of around $2800$ bits/Joule with the MMEE design, which is almost five times that of the
EE that can be achieved with the GEE-Max-based design. Similarly,
the PF-based design outperforms the GEE-Max-based design in terms of the
performance for the weakest user. However, MMEE achieves the best EE
for the weakest user compared to the other two designs. This is
because   MMEE maximizes the minimum achievable EE between all the
users and attains the same EE for all users.
 \begin{figure}[h]\label{fig:nana1}
\includegraphics[scale=.23,center]{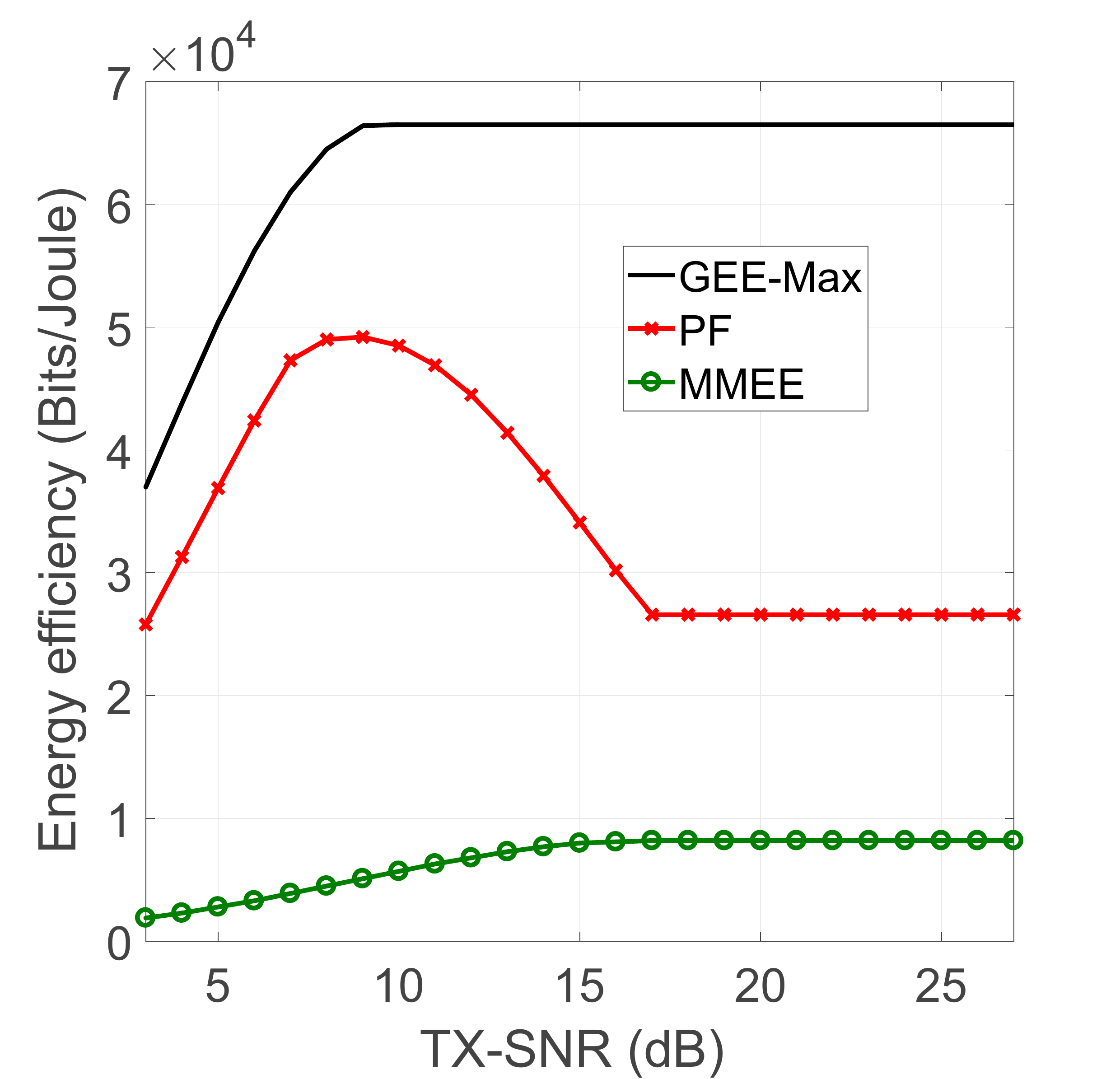}
\caption{EE of the system (i.e., GEE) with different EE-based designs.}
\centering
\label{fig:nala}
\end{figure}
Next, we compare the achieved EE  of the system (i.e., GEE) for
different designs in Fig.~\ref{fig:nala}.  As expected, the GEE-Max design outperforms the
other fairness-based designs in terms of the   EE of the system,
whereas the MMEE-based design shows the worst  GEE  performance
between the three schemes presented in Fig.~\ref{fig:nala}. However,
the PF-based design attains a good balance between the   EE at
the system level and the achieved individual EE for each user. In
other words, the PF-based design shows a better GEE
compared to that of the MMEE-based design. The same design
significantly improves EE of the weakest user compared to that of the
GEE-Max-based design.
\begin{figure}[h]\label{fig:sfig122}
\includegraphics[scale=.23,center]{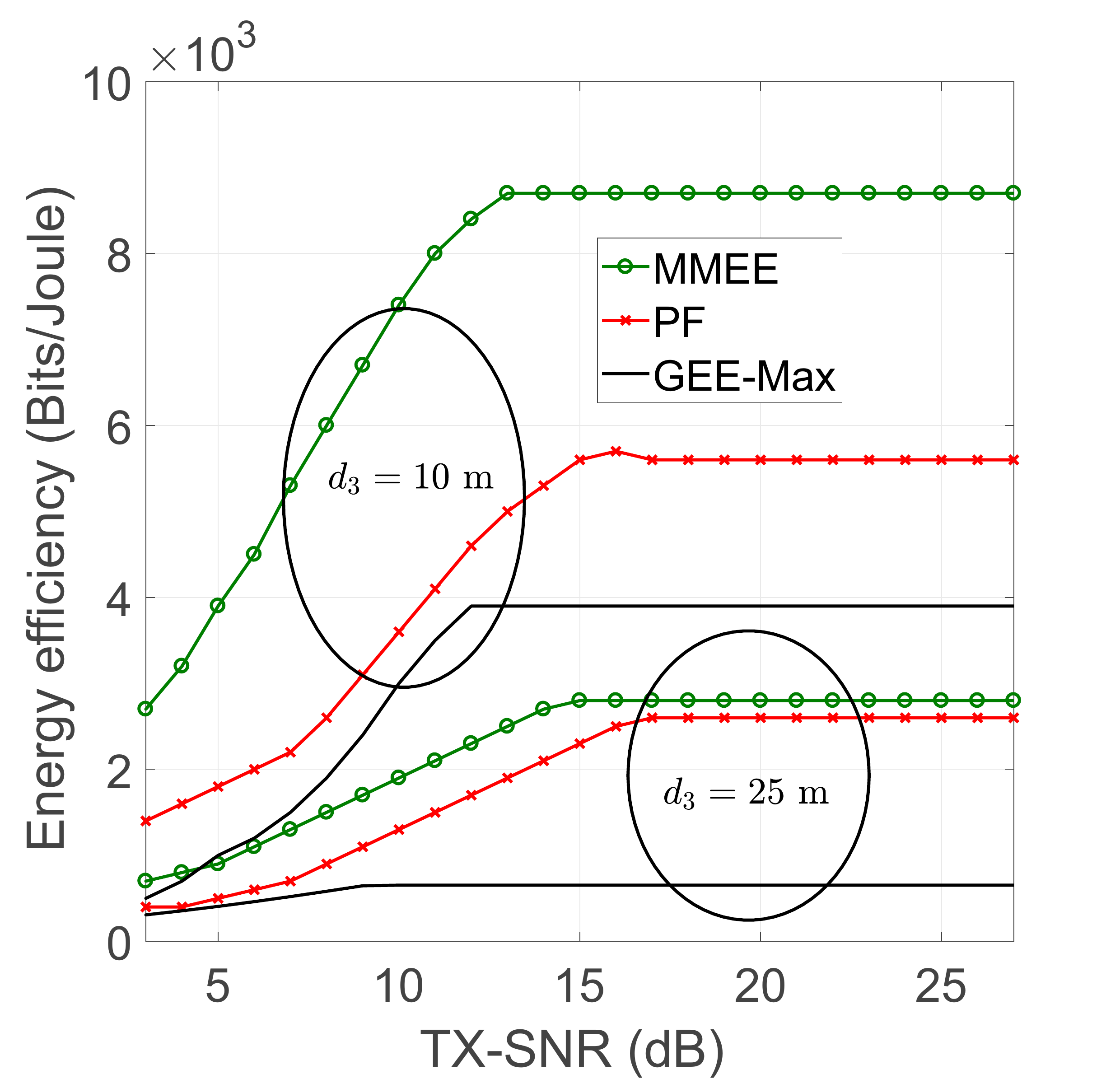}
\caption{The weakest user's performance with different designs at  different weak users' distances from the BS. }
\centering  
\label{fig:nala4} 
\end{figure}
 Finally, Fig.~\ref{fig:nala4} illustrates the effect of weakest user distance (i.e., $d_3$) on its EE performance via different beamforming designs.    As seen, the EE of the weakest user
decreases as the distance from the BS increases.

\section{Conclusions} \label{sec5}
In this paper, we proposed two beamforming designs for a MISO NOMA
system considering the EE fairness between users. The proposed designs
are based upon the MMEE and PF. In particular, we formulated the
designs as optimization problems, and applied the SCA technique  
to address the non-convexity nature of the original
problems. Furthermore,   simulation results showed  that the
MMEE-based design offers the best performance in terms of the weakest
user EE when compared to the other designs. However, this improvement
was attained at the cost of the GEE degradation of the system.
Furthermore, the PF-based design shows   a good balance between the
GEE performance   and the achieved EE of the
weakest user.
 
{  \section*{Acknowledgement}
 The work of K. Cumanan, A.  Burr and Z. Ding was supported by H2020-MSCA-RISE-2015 under grant no: 690750. }

\bibliographystyle{IEEEtran}
\bibliography{references}

\begin{thebibliography}{10}
\providecommand{\url}[1]{#1}
\csname url@samestyle\endcsname
\providecommand{\newblock}{\relax}
\providecommand{\bibinfo}[2]{#2}
\providecommand{\BIBentrySTDinterwordspacing}{\spaceskip=0pt\relax}
\providecommand{\BIBentryALTinterwordstretchfactor}{4}
\providecommand{\BIBentryALTinterwordspacing}{\spaceskip=\fontdimen2\font plus
\BIBentryALTinterwordstretchfactor\fontdimen3\font minus
  \fontdimen4\font\relax}
\providecommand{\BIBforeignlanguage}[2]{{%
\expandafter\ifx\csname l@#1\endcsname\relax
\typeout{** WARNING: IEEEtran.bst: No hyphenation pattern has been}%
\typeout{** loaded for the language `#1'. Using the pattern for}%
\typeout{** the default language instead.}%
\else
\language=\csname l@#1\endcsname
\fi
#2}}
\providecommand{\BIBdecl}{\relax}
\BIBdecl

\bibitem{fnoma}
Y.~Saito, Y.~Kishiyama, A.~Benjebbour, T.~Nakamura, A.~Li, and K.~Higuchi,
  ``Non-orthogonal multiple access {(NOMA)} for cellular future radio access,''
  in \emph{Proc. IEEE VTC Spring}, 2013, pp. 1--5.

\bibitem{noma2}
S.~Tomida and K.~Higuchi, ``Non-orthogonal access with {SIC} in cellular
  downlink for user fairness enhancement,'' in \emph{Proc. Inter. Symp. on
  Intell. Signal Process. and Comm. Systems (ISPACS)}, 2011, pp. 1--6.

\bibitem{noma4}
S.~R. Islam, N.~Avazov, O.~A. Dobre, and K.-S. Kwak, ``Power-domain
  non-orthogonal multiple access {(NOMA)} in {5G} systems: potentials and
  challenges,'' \emph{IEEE Commun. Surveys Tuts.}, 2017.

\bibitem{fayzeh}
F.~Alavi, K.~Cumanan, Z.~Ding, and A.~G. Burr, ``Beamforming techniques for
  non-orthogonal multiple access in {5G} cellular networks,'' \emph{IEEE Trans.
  Veh. Technol.}, vol.~67, no.~10, pp. 9474--9487, Oct.~2018.

\bibitem{misonoma}
Z.~Chen, Z.~Ding, P.~Xu, and X.~Dai, ``Optimal precoding for a {Q}o{S}
  optimization problem in two-user {MISO-NOMA} downlink,'' \emph{IEEE Commun.
  Lett.}, vol.~20, no.~6, pp. 1263--1266, Jun.~2016.

\bibitem{mimonoma}
Z.~Ding, F.~Adachi, and H.~V. Poor, ``The application of {MIMO} to
  non-orthogonal multiple access,'' \emph{IEEE Trans. Wireless Commun.},
  vol.~15, no.~1, pp. 537--552, Jan.~2016.

\bibitem{cuma8}
F.~Alavi, K.~Cumanan, Z.~Ding, and A.~G. Burr, ``Robust beamforming techniques
  for non-orthogonal multiple access systems with bounded channel
  uncertainties,'' \emph{IEEE Commun. Lett.}, vol.~21, no.~9, pp. 2033--2036,
  2017.

\bibitem{book}
R.~Vannithamby and S.~Talwar, \emph{Towards 5G applications: requirements and
  candidate technologies}.\hskip 1em plus 0.5em minus 0.4em\relax John Wiley \&
  Sons, 2017.

\bibitem{haitham}
H.~Alobiedollah, K.~Cumanan, J.~Thiyagalingam, A.~G. Burr, Z.~Ding, and O.~A.
  Dobre, ``Energy efficient beamforming design for {MISO} non-orthogonal
  multiple access systems,'' \emph{Accepted on IEEE Trans. Commun.}, Feb.~2019.

\bibitem{ee1}
A.~Zappone and E.~Jorswieck, ``Energy efficiency in wireless networks via
  fractional programming theory,'' \emph{Found. Trends Commun. Inf. Theory},
  vol.~11, no. 3-4, pp. 185--396, Jan.~2015.

\bibitem{fairness}
H.~Shi, R.~V. Prasad, E.~Onur, and I.~Niemegeers, ``Fairness in wireless
  networks: {I}ssues, measures and challenges,'' \emph{IEEE Commun. Surveys
  Tuts.}, vol.~16, no.~1, pp. 5--24, 2014.

\bibitem{pricefair}
D.~Bertsimas, V.~F. Farias, and N.~Trichakis, ``The price of fairness,''
  \emph{Operations Research}, vol.~59, no.~1, pp. 17--31, 2011.

\bibitem{cuma10}
K.~Cumanan, R.~Krishna, Z.~Xiong, and S.~Lambotharan, ``Sinr balancing
  technique and its comparison to semidefinite programming based qos provision
  for cognitive radios,'' in \emph{Proc. VTC Spring}.\hskip 1em plus 0.5em
  minus 0.4em\relax IEEE, 2009, pp. 1--5.

\bibitem{maxmin}
B.~Radunovi{\'c} and J.-Y.~L. Boudec, ``A unified framework for max-min and
  min-max fairness with applications,'' \emph{IEEE/ACM Trans. Netw. (TON)},
  vol.~15, no.~5, pp. 1073--1083, Oct.~2007.

\bibitem{fairmain}
F.~Kelly, ``Charging and rate control for elastic traffic,'' \emph{Trans. on
  Emerging Telecommunications Techn.}, vol.~8, no.~1, pp. 33--37, Feb.~1997.

\bibitem{hanif}
M.~F. Hanif, Z.~Ding, T.~Ratnarajah, and G.~K. Karagiannidis, ``A
  minorization-maximization method for optimizing sum rate in the downlink of
  non-orthogonal multiple access systems,'' \emph{IEEE Trans. Signal Process.},
  vol.~64, no.~1, pp. 76--88, Jan.~2016.

\bibitem{noma3}
S.~Vanka, S.~Srinivasa, Z.~Gong, P.~Vizi, K.~Stamatiou, and M.~Haenggi,
  ``Superposition coding strategies: Design and experimental evaluation,''
  \emph{IEEE Trans. Wireless Commun.}, vol.~11, no.~7, pp. 2628--2639,
  Jul.~2012.

\bibitem{cuma9}
P.~Xu, K.~Cumanan, and Z.~Yang, ``Optimal power allocation scheme for noma with
  adaptive rates and alpha-fairness,'' in \emph{Proc. IEEE GLOBECOM}, 2017, pp.
  1--6.

\bibitem{CVX}
M.~Grant, S.~Boyd, and Y.~Ye, ``{CVX}: {M}atlab software for disciplined convex
  programming,'' [Online]. Available: http://www.stanford.edu/boyd/cvx.

\bibitem{ee2}
O.~Tervo, L.-N. Tran, and M.~Juntti, ``Optimal energy-efficient transmit
  beamforming for multi-user {MISO} downlink,'' \emph{IEEE Trans. Signal
  Process.}, vol.~63, no.~20, pp. 5574--5588, Oct.~2015.

\bibitem{sqa}
A.~Beck, A.~Ben-Tal, and L.~Tetruashvili, ``A sequential parametric convex
  approximation method with applications to nonconvex truss topology design
  problems,'' \emph{J. Global Optimiz.}, vol.~47, no.~1, pp. 29--51, 2010.

\bibitem{cvx2}
Z.-Q. Luo and W.~Yu, ``An introduction to convex optimization for
  communications and signal processing,'' \emph{IEEE J. Sel. Areas Commun.},
  vol.~24, no.~8, pp. 1426--1438, Aug.~2006.

\bibitem{wcnc2}
H.~Alobiedollah, K.~Cumanan, J.~Thiyagalingam, A.~G. Burr, Z.~Ding, and O.~A.
  Dobre, ``Sum rate fairness trade-off-based resource allocation technique for
  {MISO NOMA} systems,'' in \emph{Proc. IEEE WCNC'19}.

\bibitem{cuma11}
K.~Cumanan, R.~Krishna, V.~Sharma, and S.~Lambotharan, ``Robust interference
  control techniques for multiuser cognitive radios using worst-case
  performance optimization,'' in \emph{Proc. Asilomar Conf. Signal, Syst.
  Comput}.\hskip 1em plus 0.5em minus 0.4em\relax IEEE, 2008, pp. 378--382.

\end{thebibliography}

\end{document}